\documentclass[fleqn,usenatbib]{mnras}

\usepackage{amsmath}
\usepackage{newtxtext,newtxmath}
\usepackage[T1]{fontenc}
\newcommand{\mtp}{PSR~J1710$-$3452}
\DeclareRobustCommand{\VAN}[3]{#2}
\let\VANthebibliography\thebibliography
\def\thebibliography{\DeclareRobustCommand{\VAN}[3]{##3}\VANthebibliography}

\usepackage{graphicx}	% Including figure files
\usepackage{amsmath}	% Advanced maths commands
\title[Discovery of PSR~J1710$-$3452]{Discovery of an Extremely Intermittent Periodic Radio Source}

\author[M. P. Surnis et al.]{M. P. Surnis,$^{1,2}$\thanks{E-mail: msurnis@gmail.com}
K. M. Rajwade,$^{2,3}$\thanks{E-mail: rajwade@astron.nl}
\thanks{First two authors have contributed equally to the manuscript}
B. W. Stappers,$^{2}$
G.Younes,$^{4,5}$
M. C. Bezuidenhout,$^{6}$
M. Caleb,$^{7,8}$
\newauthor
L. N. Driessen,$^{7,9}$
F.~Jankowski,$^{10, 2}$
M. Malenta,$^{2}$
V. Morello,$^{2,11}$
S. Sanidas,$^{2}$
E. Barr,$^{12}$
M. Kramer,$^{12}$
R. Fender,$^{13}$
\newauthor
and P. Woudt$^{14}$
\\
$^{1}$Department of Physics, Indian Institute of Science Education and Research Bhopal,
Bhopal Bypass Road, Bhauri, Bhopal 462 066, Madhya Pradesh, India\\
$^{2}$Jodrell Bank Centre for Astrophysics, Department of Physics and Astronomy, The University of Manchester, Manchester, M13 9PL, UK\\
$^{3}$ASTRON, the Netherlands Institute for Radio Astronomy, Oude Hoogeveensedijk 4, 7991 PD Dwingeloo, The Netherlands\\
$^{4}$ Astrophysics Science Division, NASA Goddard Space Flight
Centre, Greenbelt, 20771, Maryland, USA.\\
$^{5}$ Department of Physics, George Washington University,
Washington, 20052, DC, USA.\\
$^{6}$ Centre for Space Research, North-West University, Potchefstroom 2351, South Africa\\
$^{7}$ Sydney Institute for Astronomy, School of Physics, The University of Sydney, NSW 2006, Australia\\
$^{8}$ASTRO3D: ARC Centre of Excellence for All-sky Astrophysics in 3D, ACT 2601, Australia\\
$^{9}$CSIRO, Space and Astronomy, PO Box 1130, Bentley, WA 6102, Australia\\
$^{10}$LPC2E, Universit\'{e} d'Orl\'{e}ans, CNRS, 3A Avenue de la Recherche Scientifique, 45071 Orl\'{e}ans, France\\
$^{11}$SKA Observatory, Jodrell Bank, Lower Withington, Macclesfield, Cheshire, SK11 9FT, UK\\
$^{12}$Max-Planck-Institut f\"ur Radioastronomie, Auf dem H\"ugel 79, D-53121 Bonn, Germany\\
$^{13}$Department of Physics, Astrophysics, University of Oxford, Denys Wilkinson Building, Keble Road, Oxford OX1 3RH, UK\\
$^{14}$Department of Astronomy, University of Cape Town, Private Bag X3, Rondebosch 7701, South Africa
}

\date{Accepted XXX. Received YYY; in original form ZZZ}

\pubyear{2023}

\begin{document}
\label{firstpage}
\pagerange{\pageref{firstpage}--\pageref{lastpage}}
\maketitle

% Abstract of the paper
\begin{abstract}
We report the serendipitous discovery of an extremely intermittent radio pulsar,~\mtp, with a relatively long spin period of 10.4 s. The object was discovered through the detection of 97 bright radio pulses in only one out of 66 epochs of observations spanning almost three years. The bright pulses have allowed the source to be localised to a precision of 0.5" through radio imaging. We observed the source location with the \textit{Swift} X-ray telescope but did not detect any significant X-ray emission. We did not identify any high-energy bursts or multi-frequency counterparts for this object. The solitary epoch of detection hinders the calculation of the surface magnetic field strength, but the long period and the microstructure in the single-pulses resembles the emission of radio-loud magnetars. If this is indeed a magnetar, it is located at a relatively high Galactic latitude (2.9$^{\circ}$), making it potentially one of the oldest and the most intermittent magnetars known in the Galaxy. The very short activity window of this object is unique and may point towards a yet undetected population of long period, highly transient radio emitting neutron stars.
\end{abstract}

% Select between one and six entries from the list of approved keywords.
% Don't make up new ones.

%%%%%%%%%%%%%%%%%%%%%%%%%%%%%%%%%%%%%%%%%%%%%%%%%%

%%%%%%%%%%%%%%%%% BODY OF PAPER %%%%%%%%%%%%%%%%%%

\section{Introduction}
\label{sec:intro}

Magnetars are neutron stars with high surface magnetic field strengths of the order of $10^{14}$~G, with some exceptions) and periods ranging between 2 $-$ 12 s. Most are bright and persistent X-ray emitters with luminosities exceeding the spin-down energy. Sometimes, they exhibit luminous high-energy bursts, accompanied by erratic rotational behaviour, unlike their rotation-powered canonical neutron star cousins~\citep[see][for a review]{kb17}. This high-energy emission is believed to be powered by the decay of the ultra-strong magnetic field \citep{dt92} as the energy required for the emission is much larger than what is available from the inferred spin-down luminosity of the neutron star. Only 6 out of 26 known magnetars have shown radio emission so far (McGill magnetar catalog \citep{ok14}), which is usually observed after a high-energy outburst and fades away at later times. Given this small sample of radio-loud magnetars, the origin and mechanism of their emission is not well understood. The radio emission from the radio-loud magnetars (see for instance, cases of XTE~J1810-197 and Swift J1818.0-1607) is also highly transient in nature. It starts and stops abruptly and shows rapid variability in pulse profile shape, flux density and spectral index \citep[see][and references therein]{crh+16,mjs+19,ccc+20,erb+20,crd+21}. 

Unlike magnetars, the majority of pulsars are observed as persistent periodic radio emitters. However, their radio emission shows variability over a wide range of timescales ranging from nulling, which is the cessation of emission over a few rotation periods~\citep{wmj07} to the months to years-long disappearance of emission in intermittent pulsars \citep{klo+06,crc+12,llm+12,lsf+17}. There are a few neutron stars that show transient emission paired with magnetar-like bursts, such as PSR J1119$-$6127 and PSR J1622$-$4950. PSR J1119$-$6127 is a high magnetic field pulsar that has emitted magnetar-like bursts followed by soft X-ray flares \citep{glk+16}. On the other hand, PSR J1622$-$4950 was discovered through its pulsed radio emission and was classified as a magnetar on account of its emission at X-rays and inferred magnetic field strength. Unlike the rest of the radio-loud magnetars, for PSR J1622$-$4950, no known high-energy burst was detected before the radio emission is believed to have started \citep{lbb+10}. 

Luminous, transient radio emission from radio-loud magnetars is also of interest because of the similarities to the enigmatic Fast Radio Bursts (FRBs): microsecond to millisecond duration radio flashes that are cosmological in nature~\citep[e.g.][]{phl2022,ck21}. Recently, a Galactic magnetar, SGR J1935+2154, emitted bright radio bursts following increased high-energy activity. One of these bursts approached the luminosity of FRBs, suggesting, perhaps a connection between magnetars and FRB progenitors~\citep{brb+20, abb+20}. Transient radio emission from known and newly discovered magnetars might be the key in understanding the potential connection between FRBs and magnetars.

Here, we report the discovery of an extremely transient radio emitting neutron star (RENS), PSR J1710$-$3452. This object shows characteristics similar to radio-loud magnetars and presents an intriguing example of a long-period pulsar that straddles the boundary between regular and transient RENS. In this paper, we describe the discovery and multi-wavelength counterpart search in Section \ref{sec:obs}. The results of single-pulse analysis are descibed in Section \ref{sec:res} and the implications of our results are discussed in Section \ref{sec:disc}. Finally, we conclude our findings in Section \ref{sec:conc}.

\section{Observations}
\label{sec:obs}

\subsection{Discovery and Localisation}

\begin{figure}
    \centering
    \includegraphics[width=\columnwidth]{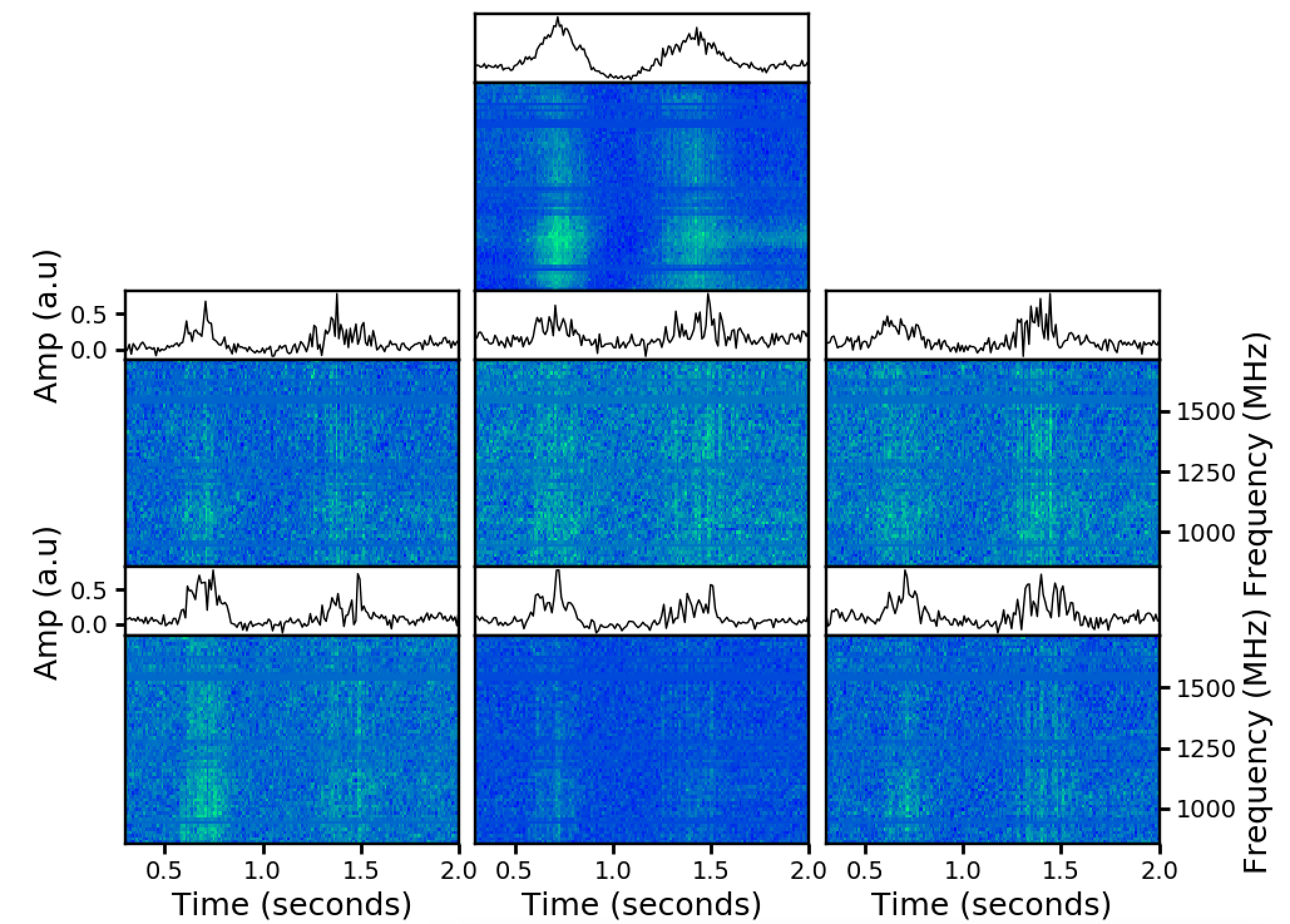}
    \caption{A gallery of some of the single radio pulses from \mtp~along with the average radio pulse profile (top most plot). For each pulse, the top panel shows intensity versus time and the bottom panel shows the dedispersed dynamic spectrum. This plot shows the diversity in the emission showing sub-structure and quasi-periodic sub-pulses. a.u. here stands for arbitrary units.}
    \label{fig:sp}
\end{figure}

PSR J1710$-$3452 was serendipitously discovered through its single-pulses on 21 June 2021 during an observation of the millisecond pulsar (MSP), PSR J1708$-$3506 with the MeerKAT telescope~\citep{jonas2016}, as part of the MeerTIME program \citep{bja+20}. The single-pulses were detected through the Meer(more) TRansients and Pulsars (MeerTRAP) pipeline which is a real-time commensal mode back-end on the MeerKAT telescope~\citep{rbc+22}. The observation lasted 21 minutes and resulted in the detection of 97 pulses (see Figure \ref{fig:sp} for examples) from the object with a dispersion measure (DM) of 189 $\pm$ 0.5  pc cm$^{-3}$. A simple periodicity search resulted in the clear detection of a period of 10.4~seconds between the radio pulses. The pulses were detected in the incoherent beam with a total field of view of $\sim$ 1.3 sq. degree and hence, the localisation of the new source was poor. A search for known sources within the incoherent beam revealed the bright 0.7 s pulsar, PSR~J1708$-$3426 with a similar DM (190 $\pm$ 2~pc~cm$^{-3}$). While the DM of PSR~J1708-3426 was within the 1-$\sigma$ uncertainty range of the DM of \mtp, the measured period was not harmonically related to the period of PSR~J1708$-$3426, leading us to believe that \mtp~was a new object.  

The only way to conclusively prove \mtp~as a newly discovered object was to localise it in the radio images. The signal to noise ratio (S/N) of the single-pulses from \mtp~was high enough for it to be detected in the synthesis radio images made from the visibilities recorded simultaneously. The minimum integration time of 8~s for the visibilities compared to the rotation period of 10.4~s resulted in some integrations without any pulsed emission. Using the known period and the time-domain data, we made single `ON' and `OFF' pulse images by adding an equal number of `ON' and `OFF' pulse integrations. Subtracting the two images resulted in the detection of the source at the sky location RA 17:10:22.81(3), DEC $-$34:52:57.5(5). This position also confirmed that it was not associated with PSR~J1708$-$3426, which is located more than 33' away.

\begin{figure}
    \centering
    \includegraphics[width=\linewidth]{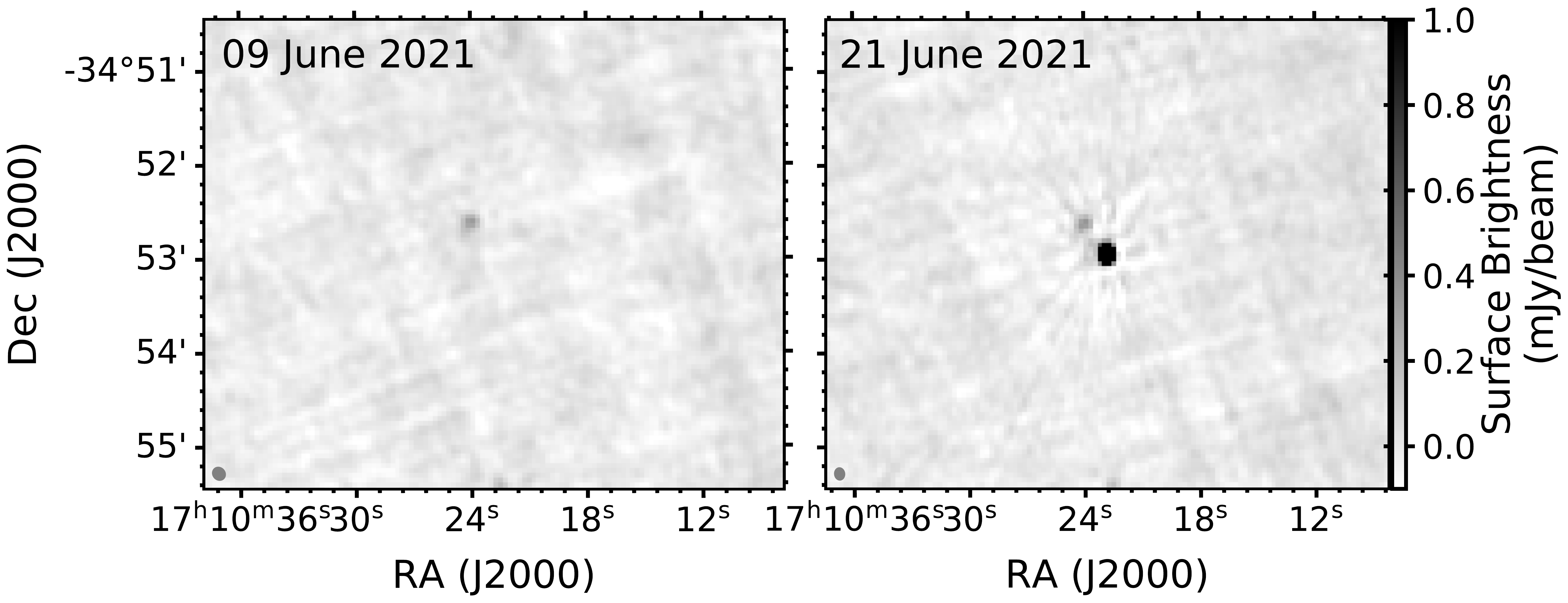}
    \caption{MeerKAT images of the position of PSR J1710$-$3452 on 09 June 2021 (left) and 21 June 2021 (right). The synthesised beam is shown as a grey circle in the bottom left corner of each panel. PSR J1710$-$3452 can be clearly seen in the right hand image.}
    \label{fig:onoff}
\end{figure}

Using an image made from the entire 21 minute observation (right panel of Figure \ref{fig:onoff}), we determined a flux density for the continuum source at the position of \mtp~of 12.5 $\pm$ 1.3 mJy at 1283.6 MHz\footnote{The beam correction is approximate for the MTMFS algorithm, so we have put a 10\% uncertainty to be conservative}. As the nearby MSP PSR J1708$-$3506 is part of the monitoring campaign led by the MeerTIME project~\citep{bja+20}, the field is being observed regularly with a cadence of roughly two weeks. At the time of writing the paper, this field had been observed for total of 66 epochs, totalling 1320 minutes of observing time. Although no previous detections of bright single-pulses had been made by MeerTRAP, we investigated whether the source might have been present previously, but either fainter, or emitting in a different mode. We imaged the MeerKAT interferometric data from a total of 19 epochs between 9 June 2021 and 13 July 2022. All the radio images except the discovery epoch resulted in non-detection of the object with the median and the most stringent 3-$\sigma$ upper limits on the flux density being 81 and 51 $\mu$Jy, respectively (see Figure \ref{fig:onoff} for an example). This puts the flux density ratio between the `ON' and `OFF' states to be greater than 240 for the most stringent limit. Since there was a direct correlation between non-detections in time domain and radio images, we were confident that if no pulses were detected by the MeerTRAP back-end then the source was not emitting radio pulses. We searched through archival radio images from different telescopes to check if the continuum source was detected elsewhere. These included the TIFR-GMRT Sky Survey (TGSS) \citep{ijm+17} at 147.5 MHz, the Rapid ASKAP Continuum Survey (RACS) \citep{mhl+20} at 887.5 MHz and the NRAO-VLA Sky Survey (NVSS) \citep{ccg+98} at 1400 MHz. We did not detect the continuum source in any of the radio images with 3-$\sigma$ flux density upper limits of 60, 1.2 and 1.95 mJy at 147.5, 887.5 and 1400 MHz, respectively. These survey observations encompass the duration of almost 26 years starting with the NVSS observation which was carried out on 17 January 1995\footnote{Exact dates were not available for the TGSS (between 2010$-$2012) and RACS (between 2019$-$2021) observations.}.

\subsection{Multi-wavelength Counterpart}

In order to check if the detection of radio pulses was associated with the onset of any high-energy emission, we looked for real-time triggers from three all-sky burst monitors: \textit{Swift-BAT}, \textit{Fermi-GBM} and \textit{INTEGRAL}~\citep{krimm2013, meegan2009,winkler2003}. We searched for triggers issued by the respective burst report systems from the direction of \mtp~for up to two weeks prior to the detection of radio pulses. Along with reported bursts, we also checked for sub-threshold bursts in GBM, i.e bursts that are just below the detection threshold of the instrument~\citep[see][for more details]{younes20ApJ}. No confirmed bursts were found in this search although a few sub-threshold burst candidates that were nominally located towards \mtp~were found in the \textit{Fermi-GBM} data. For these bursts, the localisation error was too large to confirm any association. We also searched for archival data on this field in observations with the \textit{Swift-XRT}, \textit{XMM-Newton} and \textit{Chandra} X-ray telescopes but unfortunately none of these telescopes have observed this field. Consequently, we carried-out Target of Opportunity observations with the \textit{Swift}-XRT telescope for a total of 2.5~ks in the Photon Counting mode. We did not detect an X-ray source at the location of~\mtp~and place a 3-$\sigma$ upper limit on the absorbed X-ray flux of 4.3$\times$10$^{-13}$~ergs~cm$^{-2}$~s$^{-1}$. The measured limit is not very constraining for emission models and deeper X-ray observations with more sensitive telescopes like \textit{XMM-Newton} or \textit{Chandra} are needed to reveal if \mtp\, is an X-ray emitter.

We also checked the position of \mtp~in all the archival optical surveys, namely the Digitized Sky Survey (DSS)~\footnote{\url{http://gsss.stsci.edu/zzzOldWebSite/DSS/dss_home.htm}} and the Dark energy CAmera Plane Survey (DeCAPS)~\citep{sgl+17}. We did not detect any optical source at the location of~\mtp. While we see a hint of emission in the \textit{z}-band image from DeCAPS, it is not significant ($\sim$1.5--2~$\sigma$) enough to warrant further investigation. From the non-detections, we calculate 5-$\sigma$ upper limits of 23.5, 22.6, 22.1, 21.6 and 20.3 magnitude in the \textit{grizY} bands from~\cite{ssl+22}. We also looked for an infra-red counterpart in the 2-Micron All Sky Survey ~\citep[2MASS;][]{scs+06} and did not find any. The 10-$\sigma$ upper limits on the infra-red magnitudes in the \textit{J}, \textit{H} and \textit{K$_{s}$} bands are 15.8, 15.1 and 14.3 respectively.

\section{Results}
\label{sec:res}

\subsection{Timing}

\begin{figure}
    \centering
    \includegraphics[width=\linewidth]{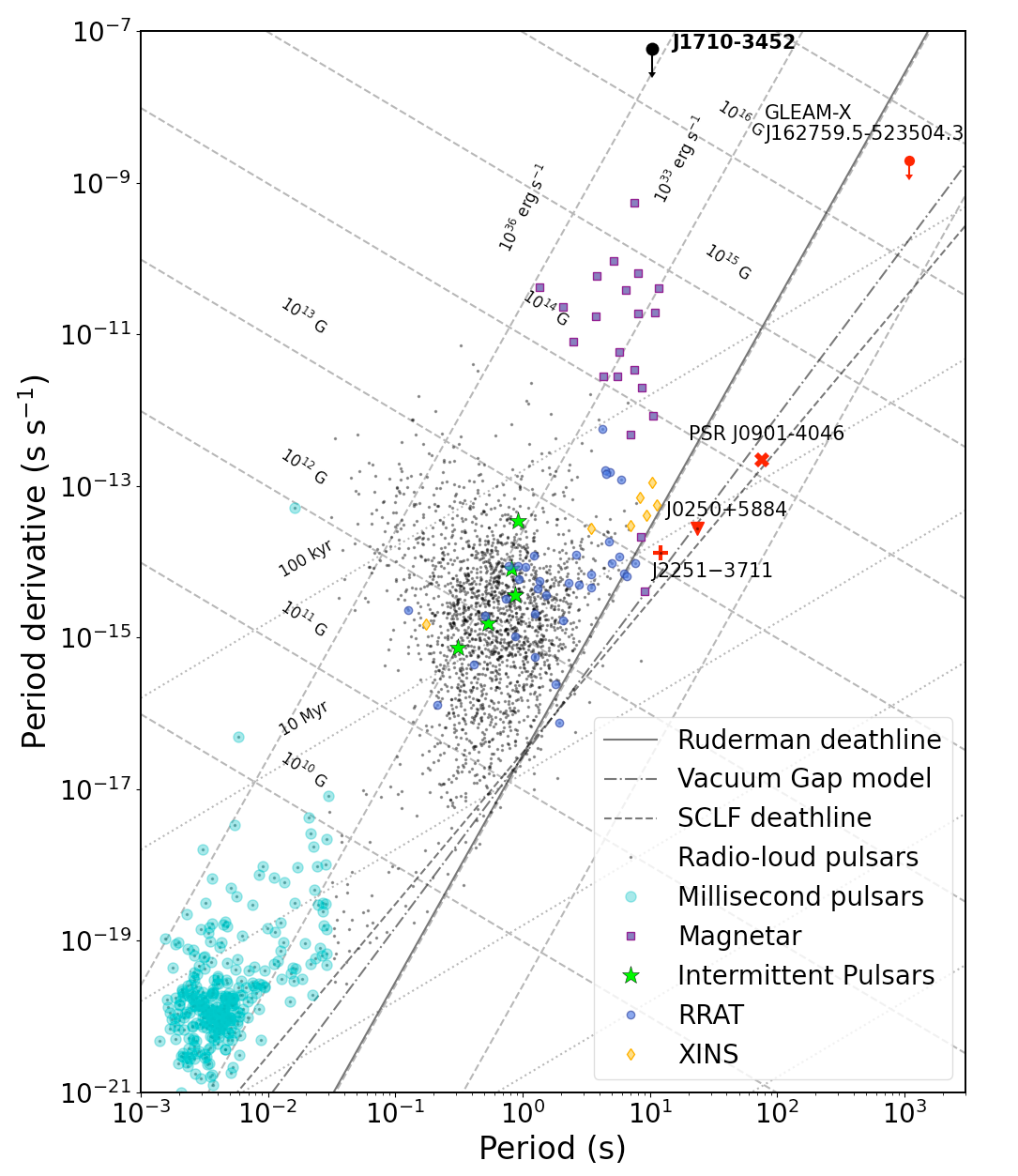}
    \caption{Period-Period derivative plot showing the known pulsar population (ATNF pulsar catalog version 1.69). In particular we highlight the X-ray Isolated Neutron Stars (XINS), Magnetars, and Rotating Radio transients (RRATs). The large black circle with an upper limit arrow shows the best constrained position of~\mtp~in the diagram. Various representative pulsar death-lines from different theoretical models are also shown.}
    \label{fig:ppdot}
\end{figure}

We undertook a timing analysis using the single-pulses and derived a rotation period of 10.412(1)\,s. Our single epoch detection allowed us to only obtain an upper limit of 5.9$\times$10$^{-8}$ $s s^{-1}$ for the period-derivative, which corresponds to an upper limit of 2.5$\times$10$^{16}\mbox{ } G$ on the surface magnetic field strength. This places \mtp ~in the region of the $P-\dot{P}$ diagram that is occupied by magnetars and high magnetic field pulsars (see Figure \ref{fig:ppdot}) or close to the XINS or below the pulsar death lines if we assume that~\mtp~comes from an older population of magnetars (see Section 4.3).

\subsection{Pulse Shape and Microstructure}

\begin{figure*}
    \centering
    \includegraphics[width=\textwidth]{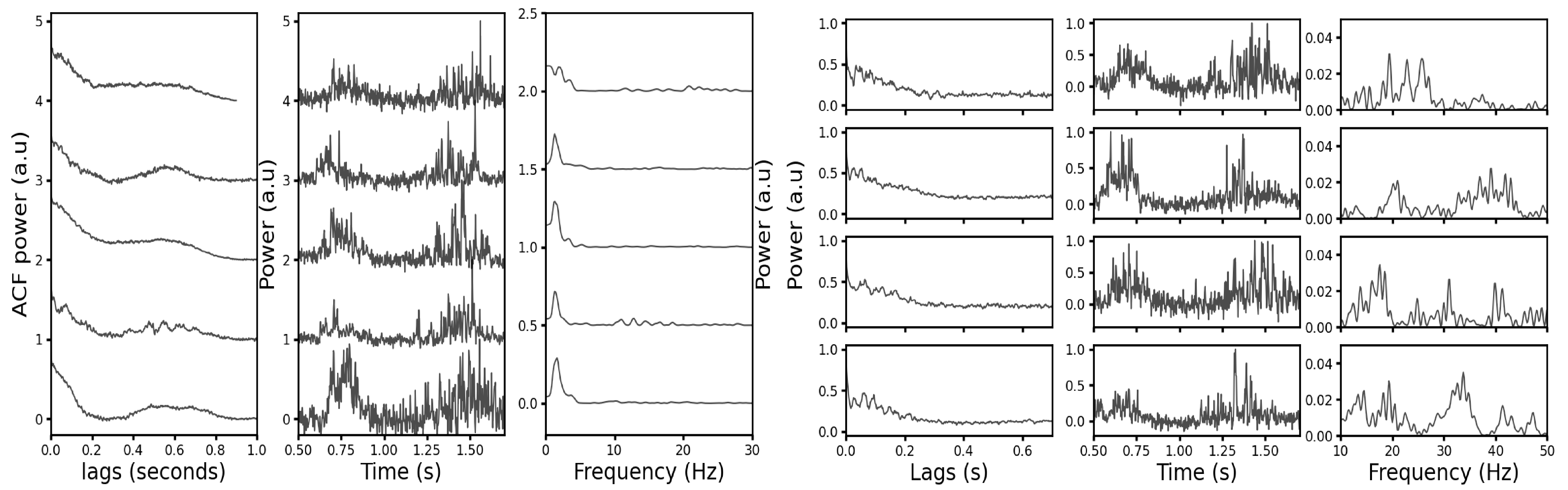}
    \caption{\textit{Left:} Examples of single-pulses from \mtp~showing the sub-pulse structure (middle) with the ACF (left) and the Lomb-Scargle periodogram (right). \textit{Right:} Four examples of single-pulses showing quasi-periodic structure (middle) with the ACF of the trailing component (left) and the corresponding Lomb-Scargle periodogram (right). Here, a.u stands for arbitrary units.}
    \label{fig:sp_analysis}
\end{figure*}

The single-pulses show broadly similar emission but there is a large diversity of structure on a smaller timescale. Every single-pulse shows two main emission components; each comprising several narrow, broadband and in some instances, quasi-periodic sub-pulses (see Figure \ref{fig:sp_analysis} for example). The two main components are separated by approximately 0.7~seconds with a total pulse width spanning more than 1~second, corresponding to a pulsed duty cycle of $\sim$10$\%$. This large duty cycle does not follow the typical duty-cycle-period anti-correlation seen in RENS~\citep{jk19} and is closer to what is observed in radio-loud magnetars~\citep{crd+21}. To quantify the narrow sub-structure, we computed the auto-correlation function (ACF) for each single-pulse. For total intensity at each sample, $I_{t}$, the ACF is given by:
\begin{equation}\label{eq:acf}
    \text{ACF}(\Delta t) = \frac{\sum_{t}(I(t))(I(t+\Delta t)) }{\sqrt{\sum_{t}(I(t))^2  \sum_{t}(I(t+\Delta t))^2}},
\end{equation}
where $I(t + \Delta t)$ is the intensity of the sample at a delay of $\Delta t$. We computed the ACF for a range of $\Delta t$ in order to see any systematic structure of the single-pulses. For pulses where quasi-periodic microstructure was clearly visible, we measured the distance between the minima and maxima of the different peaks (see right panel of figure~\ref{fig:sp_analysis}). These measurements lie in the range 0.03-0.06 seconds. For each single-pulse, we also computed the Lomb-Scargle periodogram~\citep{scargle1982} in order to study the quasi-periodicity between the sub-bursts.

The result of this analysis for five sample pulses is shown in Figure~\ref{fig:sp_analysis}. The periodogram is dominated by the 1.35 Hz periodicity which corresponds to the separation between the two main components. However, we find quasi-periodic signals in the trailing component of the radio profile for a few single-pulses.

\section{Discussion}
\label{sec:disc}

\subsection{Transient nature of the radio emission}

Radio emission from the known sample of RENS shows a lot of diversity. One of its extreme manifestations is intermittency. Following a period of radio silence, known intermittent RENS show regular emission over a time span of months in their active window \citep{crc+12,llm+12,lsf+17}. However, there are a few exceptional examples. PSR J1832$+$0029 was only detected once in a 10 minute observation in its `OFF' state \citep{wwh+20}, while PSR J1841$-$0500 has exhibited nulls over a long duration when it was in its `ON' state \citep{crc+12}. Given that the spin-down rate of these pulsars remained unchanged during these brief emission/nulling episodes, these were possibly one-off events. PSR J1929$+$1357 provides a much more interesting case of intermittency over long `ON' and `OFF' emission states. This pulsar was initially detected in only five epochs over 100 hours of telescope time. However, the likelihood of detection seemed to increase with time. Eventually, it was seen to have a longer duty cycle as its spin-down rate increased \citep[see Figure 6 in][]{lsf+17}, which points towards magnetospheric changes as the catalyst for the modulation of the `ON' duty cycle. Given the regular nature of the observations both before and after the detection of~\mtp, any activity cycle similar to the other intermittent pulsars would have resulted in a few detections. A complete lack of detection in all the other epochs indicates a more extreme activity cycle which seems to be unique so far. This is analogous to other transient periodic sources like GLEAM$-$X J162759.5$-$523504.3 that was detected for 3 months with no detections before or after~\citep{hzb+22}, indicating a strong observational bias against this source population that may be lurking in the Galaxy in large numbers~\citep{bwh+22}. In addition, the rotation period of 10.4 s puts \mtp~ in a very different part of the $P-\dot{P}$ diagram compared to the intermittent pulsars. This part is occupied by a population of magnetars and other intermittent radio emitters like RRATs. The long spin period combined with single-pulses that show quasi-periodic sub-bursts might imply that \mtp~ belongs to a new category of long-period RENS like PSR J0901$-$4046 \citep{chr+22}, which have not been observed before due to a severe selection bias. On the other hand, radio-loud magnetars show highly transient radio emission, which could last a month to a few years \citep{lbb+10,crh+16}, yet not necessarily triggered by a high-energy outburst. A recent example is the $<1$~month radio-active episode of SGR 1935+2154 \citep{z20atel1935}, seemingly triggered by a timing anomaly \citep{y23na1935}. Hence, \mtp~ might belong to a class of such magnetars that show transient radio emission not triggered by a high-energy event. 

\subsection{Pulse sub-structure}

A small fraction of RENS exhibit sub-bursts in their single-pulses. This microstructure provides insights into the mechanism of the coherent radio emission. Radio-loud magnetars show a diverse range of emission features at radio wavelengths. These are characterised by a long emission envelope (ranging from few tens to a few 100~ms) with a large variability in every rotation. For example, pulses from the radio-loud magnetar XTE~J1810$-$197 show bright, narrow giant pulses and sub-bursts that are quasi-periodic in nature~\citep{mjs+19,crd+21,ljs+21}. The radio pulses from \mtp~share a lot of traits with radio-loud magnetars. The duty cycle is 10$\%$ and the pulses are highly variable and made up of two main components (PSR~J1622-4950; \citealt{lbb+12}). The trailing component comprises several shorter bursts with pulse widths ranging from 10--80~ms (XTE~J1810-197; \citealt{msj+22}). The ACF and the periodogram show more structure at timescales between 40--60~ms (XTE~J1810-197; \citealt{crd+21}). All these characteristics along with the rotation period indicates that \mtp\ might be a magnetar.

\subsection{Old Magnetar?}

Magnetars are typically a younger class of neutron stars (age $\sim$10~kyr) compared to the general population~\citep{kb17}. Hence, it is expected that most of the magnetars are formed and will still be found in the Galactic plane, which matches the current observations.~\mtp~has a relatively high Galactic latitude of 2.91 degrees. This would suggest that if \mtp~is a magnetar, it is much older when compared to the rest of the population. The lack of any observational evidence for a supernova remnant in the vicinity of~\mtp~supports this conjecture. Intriguingly, there is only one other magnetar that is above the Galactic plane namely, SGR 0418$+$5729. It is also the only magnetar that possesses a low surface magnetic field strength (10$^{12}$~G) but still emits magnetar-like bursts~\citep{RET+2010}. It has been suggested that SGR 0418$+$5729 is an old magnetar~\citep{rip+2013}, which can explain the low magnetic field and the low surface temperature. If \mtp~is also an old magnetar, it could explain the intermittency of the radio emission and its position in the Galaxy. Using this hypothesis, we can get an upper limit on the age of \mtp~. At a distance of 4~kpc,~\mtp~lies about 200~pc above the Galactic plane. Using the mean kick velocity of neutron stars of 150~km~s$^{-1}$~\citep{hdlk2005}, the limit on the age is $\leq$2~Myr. If we take into account the uncertainty on the distance due to the electron density models (typically a factor of 2) and a range of kick velocities from 60--540~km~s$^{-1}$~\citep{vic17}, we obtain an age range anywhere between 184~kyr and 6.6~Myr. These estimates would still make \mtp~one of the oldest magnetars to be discovered.

This discovery also begs the question whether a large population of these sources exists in the Galaxy. While the discovery of a single object makes a quantification challenging, we estimate an upper-limit on the rate of such events with the following assumptions: 1) an active episode with multiple bursts is a single event, 2) since MeerTRAP started observing, such objects have only gone through at least one such episode and 3) all the MeerTRAP pointings are observed for longer than the typical episode duration. Assuming that such events follow a Poisson distribution across the sky, the 99$\%$ confidence upper limit on the number of events following~\cite{g86} is $\sim$7. Given that the MeerTRAP incoherent beam covers approximately 1.3 square degrees at L-band and assuming that MeerTRAP has spent about 400 days on sky until July of 2022, we obtain an upper limit of $\sim$560 events per day per sky. We would like to caution the reader that this rate estimate is a very crude upper limit and a more realistic rate estimate could only be done after a significant number of such sources is discovered in the future.

\section{Conclusion}
\label{sec:conc}

We have discovered an extremely intermittent radio-emitting neutron star with a rotation period of 10.4 seconds, which we have localised through radio images. The object was detected in only 1 out of 66 epochs of observations spread over three years suggesting a very narrow activity cycle. The single-pulses show emission properties reminiscent of radio-loud magnetars but the single epoch detection hinders definitively determining that it has the expected magnetic field strength. No counterpart was detected at any other wavelength. The highly intermittent nature and its position in the Galaxy suggests that \mtp~is part of an older population of magnetars which makes it extremely interesting for evolutionary models of neutron stars. The apparent lack of emission over very long timescales with an extremely narrow activity cycle indicates the presence of a Galactic population of such sources which can be uncovered only through repeated large area sky surveys. It also highlights the potential of image plane searches to find these sources.

\section*{Acknowledgements}
 
We thank the anonymous referee whose suggestions have improved the manuscript. The MeerTRAP collaboration acknowledges funding from the European Research Council (ERC) under the European Union's Horizon 2020 research and innovation programme (grant agreement No 694745). K.M.R. acknowledges support from the Vici research program `ARGO' with project number 639.043.815, financed by the Dutch Research Council (NWO). M.C. acknowledges support of an Australian Research Council Discovery Early Career Research Award (project number DE220100819) funded by the Australian Government and the Australian Research Council Centre of Excellence for All Sky Astrophysics in 3 Dimensions (ASTRO 3D), through project number CE170100013. The MeerKAT telescope is operated by the South African Radio Astronomy Observatory, which is a facility of the National Research Foundation, an agency of the Department of Science and Innovation.

%%%%%%%%%%%%%%%%%%%%%%%%%%%%%%%%%%%%%%%%%%%%%%%%%%
\section*{Data Availability}

The data used in this work will be shared with interested parties on reasonable request to the authors.

%%%%%%%%%%%%%%%%%%%% REFERENCES %%%%%%%%%%%%%%%%%%

% The best way to enter references is to use BibTeX:

\bibliographystyle{mnras}
\bibliography{refs} % if your bibtex file is called example.bib

% Alternatively you could enter them by hand, like this:
% This method is tedious and prone to error if you have lots of references
%\begin{thebibliography}{99}
%\bibitem[\protect\citeauthoryear{Author}{2012}]{Author2012}
%Author A.~N., 2013, Journal of Improbable Astronomy, 1, 1
%\bibitem[\protect\citeauthoryear{Others}{2013}]{Others2013}
%Others S., 2012, Journal of Interesting Stuff, 17, 198
%\end{thebibliography}

%%%%%%%%%%%%%%%%%%%%%%%%%%%%%%%%%%%%%%%%%%%%%%%%%%

%%%%%%%%%%%%%%%%% APPENDICES %%%%%%%%%%%%%%%%%%%%%

\appendix

%%%%%%%%%%%%%%%%%%%%%%%%%%%%%%%%%%%%%%%%%%%%%%%%%%

% Don't change these lines
\bsp	% typesetting comment
\label{lastpage}
\end{document}